\documentclass[twocolumn,prl]{revtex4}

\usepackage{graphicx}    
\usepackage{dcolumn}     
\usepackage{bm}          

\begin{document}
\preprint{Organic}

\title{Tomography of pairing symmetry from magnetotunneling
spectroscopy --- 
a case study for quasi-1D organic superconductors 
}

\author{
Y. Tanuma$^1$, K. Kuroki$^2$, Y. Tanaka$^3$, R. Arita$^4$,
S. Kashiwaya$^5$, and H. Aoki$^4$
}%
%
\affiliation{
$^1$Graduate School of Natural Science and Technology,
Okayama University, Okayama 700-8530, Japan \\
$^2$Department of Applied Physics and
Chemistry, The University of Electro-Communications,
Chofu, Tokyo 182-8585, Japan \\
$^3$Department of Applied Physics, 
Nagoya University, Nagoya, 464-8603, Japan \\
$^4$Department of Physics,
University of Tokyo, Hongo, Tokyo 113-0033, Japan \\
$^5$National Institute of Advanced Industrial Science
and Technology, Tsukuba, 305-8568, Japan
}
%
\date{\today}
\begin{abstract}
We propose that anisotropic $p$-, $d$-, or $f$-wave pairing symmetries 
can be distinguished from a tunneling spectroscopy in the presence of 
magnetic fields, which is exemplified here for a  model 
organic superconductor $\mbox{(TMTSF)}_{2}X$.   
The shape of the Fermi surface (quasi-one-dimensional 
in this example) affects sensitively the 
pairing symmetry, which in turn affects the shape (U or V) 
of the gap along with the presence/absence of 
the zero-bias peak in the tunneling in a subtle manner.  
Yet, an application of a magnetic field enables 
us to identify the symmetry, which is interpreted as 
an effect of the Doppler shift in Andreev bound states. 
\end{abstract}
\maketitle

One of the most fascinating features of
unconventional superconductors with anisotropic 
pairing symmetries is that
Andreev reflection at surfaces can occur in a variety of ways
\cite{Buch,Hu}.
Namely, at surfaces and interfaces, 
interference takes place between
incident and reflected quasiparticles,  
which, when the pairing is anisotropic, 
experience opposite signs of the pair potential depending 
on the situation, since an anisotropic pairing 
dictates that the BCS gap function has to have node(s).  
The interference then becomes constructive, 
and we end up with the Andreev bound states (ABS)
at the surface, which should be 
observed as a zero-bias conductance peak (ZBCP)
in tunneling spectroscopy.
This is recognized  as a clear signature of an 
anisotropic pairing \cite{TK95,FRS97,Honer98,Boss}.
The ZBCP is observed
in various anisotropic superconductors 
such as the high-$T_C$ cuprates 
\cite{TK95,GXL88,Covin97,Alff,Wei}, 
another oxide $\mbox{Sr}_2\mbox{RuO}_{4}$ \cite{Laube,Mao},
and a heavy fermion system $\mbox{UBe}_{13}$ \cite{Walti}.
In this context, it is intriguing to
investigate whether the ZBCP can be observed
in organic superconductors,
another class of candidates for anisotropic pairing,
such as Bechgaard salts $\mbox{(TMTSF)}_{2}X$
($X=\mbox{PF}_{6}$, $\mbox{ClO}_{4}$, etc.)
\cite{Jerome,Bechgaard}.
\par
For $\mbox{(TMTSF)}_{2}X$, a spin-triplet pairing 
has been suggested from 
an observation of a large $H_{c2}$ \cite{Lee01} and an
unchanged Knight shift across $T_c$ \cite{Lee02}.
Anisotropic pairing with nodes on the Fermi surface 
has been suggested from an NMR measurement
\cite{Takigawa}, while a thermal conductivity
measurement has reported
the absence of nodes on the Fermi surface
\cite{BB97}.
Theoretically, an anisotropic $p$-wave pairing in which
the nodes of the pair potential can be made to avoid intersecting 
the (quasi-1D) Fermi surface has been proposed in an early stage
\cite{Abrikosov,HF87,Lebed}.
The triplet pairing itself, however, is puzzling 
in an electron mechanism for superconductivity, 
since the usual wisdom dictates that a triplet pairing should 
be an outcome of ferromagnetic spin-fluctuation exchange, whereas 
the superconductivity lies right next to a $2k_{\rm F}$ SDW
in the pressure-temperature phase diagram for TMTSF \cite{GE80}.
--- naively, SDW spin fluctuations should favor
spin-singlet $d$-wave-like pairing \cite{Shima01,KA99,KK99}.
In order to solve this puzzle, three of the present authors 
have recently proposed \cite{KAA01} that the correct pairing may be 
a triplet $f$-wave, which can 
dominate over $d$- and $p$-waves if we introduce 
$2k_{\rm F}$ CDW coexisting with the $2k_{\rm F}$ SDW, 
and a magnetic anisotropy in spin fluctuations
in this quasi-1D material.
\par
Motivated from the above mentioned 
experimental as well as theoretical controversies, 
we ask ourselves: can we identify the 
pairing symmetry ($p,d,f$) from the tunneling spectroscopy 
in general, and from the ZBCP in particular?\cite{Arai}
Recently, it has been pointed out that
these pairing symmetries in $(\mbox{TMTSF})_{2}X$
can be distinguished from the presence (for $p$ and $f$) 
or absence ($d$) of 
the ABS on the surface \cite{Sengupta}.
Subsequently we have pointed out \cite{Tanu02} that $p$ and
$f$ can still be distinguished by looking at 
the shape of the gap in the surface density of states
in which the zero-energy peak (ZEP) \cite{ZEP} resides \cite{Melo}.
\par
In real materials, however, the question is what will happen 
for a general shape of the Fermi surface.  
In $(\mbox{TMTSF})_{2}X$, the warping of the quasi-1D Fermi surface is 
actually controlled by the ratio between the transfer within 
the chains and several types of transfers across the chains.
So, if the tunneling spectrum is sensitive 
to the warping, identification of
the pairing symmetry will be marred,
so we will have to devise some {\it in situ} way of
probing the symmetry
from the tunneling spectroscopy.  
In the present study, we first show that the tunneling 
spectrum is in fact sensitive to the warping. 
This is due to the degradation of the 
symmetry in the Fermi surface against $k_a\rightarrow -k_a$
in $k$ space, and the appearance/disappearance of
the ZEP may even be inverted between $d$ and $f$-waves 
by a small change in the hopping integral.
\par
We then show that we can overcome this difficulty 
by applying a magnetic field as an {\it in situ} control.
In a magnetic field
screening current affects the ABS spectrum, 
which is called the Doppler shift,
and this can split
the ZBCP \cite{FRS97,Covin97,Aprili,Krupke}. 
We find that ZBCP splits into two for $d$-wave,
while it does not for $p$ and $f$.
The way in which the Doppler shift occurs 
reflects the shape of the Fermi surface, so the 
magnetotunneling spectrum provides 
a unique kind of tomography of the gap function
symmetry on the Fermi surface.
\par
%
\begin{figure}[htb]
\begin{center}
\scalebox{0.28}{
\includegraphics{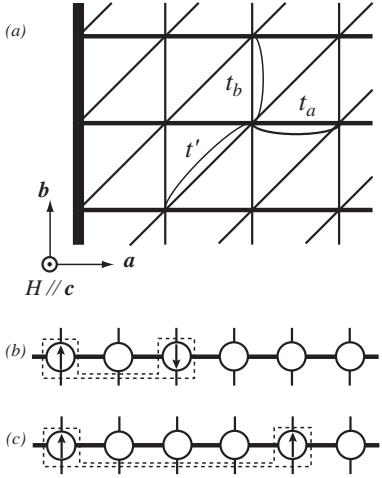}}
\caption{
(a) 
The hopping integrals
and the surface normal to the $a$-axis are shown.
Cooper pairs between sites
separated by $m_l$ lattice spacings
are illustrated
for 
(b) triplet $p$- or
singlet $d$-wave ($m_{p}=m_{d}=2$)
and (c) triplet $f$-wave ($m_{f}=4$).
\label{fig:01}
}
\end{center}
\end{figure}
Three of the present authors have derived
in Ref.~\onlinecite{KAA01}
plausible pairing symmetries by considering pairing interactions 
arising from spin and charge fluctuations.
For discussing the tunneling spectroscopy
for a given pairing symmetry, it is more convenient 
to start from the extended Hubbard model
that incorporates effective
attractions for the given pairing symmetry.
So we have
\begin{eqnarray}
\label{Hami01}
    {\mathcal H} &=&
 - \sum_{\langle \bm{i,j} \rangle,\sigma}t_{\bm{ij}}
  c^{\dagger}_{\bm{i}\sigma} c_{\bm{j}\sigma}
 - \sum_{\bm{i}\sigma} \mu c_{\bm{i}\sigma}^{\dagger} c_{\bm{i}\sigma}
\nonumber \\
 &&- \frac{V}{2}\sum_{|\bm{i}-\bm{j}|=m_l,\sigma,\sigma^{\prime}}
      c^{\dagger}_{\bm{i}\sigma}
      c^{\dagger}_{\bm{j}\sigma^{\prime}}
      c_{\bm{j}\sigma^{\prime}}
      c_{\bm{i}\sigma},
\end{eqnarray}
where $c_{\bm{i}\sigma}^{\dagger}$
creates a hole with spin $\sigma=\uparrow,\downarrow$
at site $\bm{i}=(i_{a},i_{b})$.
$V$ is the effective attraction that is assumed to act 
on a pair of electrons when they are $m_l$ lattice spacing 
apart (along $a$ axis when the system is quasi-1D), 
where  $m_l=2(4)$ for
$l=p,d(f)$-wave pairing [see Fig.~\ref{fig:01}(b)(c)].
\par
\begin{figure}[htb]
\begin{center}
\scalebox{0.33}{
\includegraphics{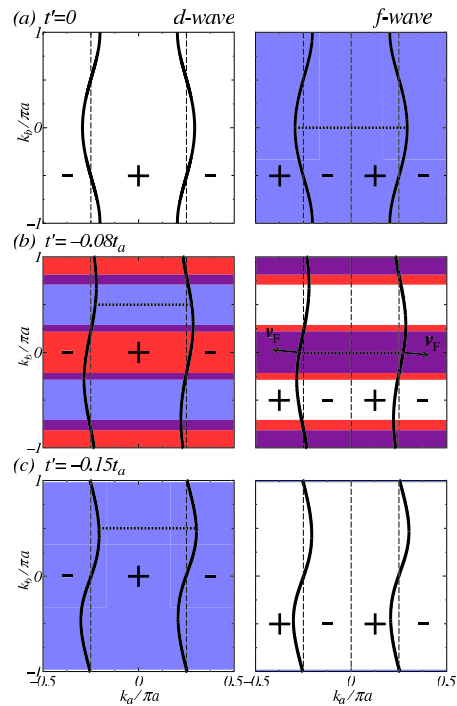}}
\caption{
Fermi surface fixed in $t_{b}/t_{a}=0.1$:
(a) $t^{\prime}=0$,
(b) $t^{\prime}/t_{a}=-0.08$,
and
(c) $t^{\prime}/t_{a}=-0.15$.
The $d$-wave and $f$-wave pair potential are
shown in the left and the right panels, respectively.
$+$ and $-$ denote the sign of the pair potential,
and the dashed lines represent the nodal line.
The blue areas indicate the region where the gap function has 
opposite signs at the ends of the dotted lines, 
so the states at the ends of the dotted lines correspond to
injected and reflected states of the quasiparticles that 
lead to Andreev bound states and thus zero-bias peaks.
In (b), the red areas indicate the region where $v_{{\rm F}b}$
has opposite signs for injected and reflected states.
\label{fig:03}}
\end{center}
\end{figure}

As for the lattice we take here an anisotropic 
($t_{b}\ll t_{a}$) 2D system.
To warp the shape of the Fermi surface we introduce 
a diagonal transfer $t'$, so the system 
is an anisotropic triangular lattice.  
The sample edge is assumed to be $\perp a$ axis.
In what follows we vary 
$t'(\leq 0)$ in the range $0\leq |t^{\prime}/t_a| \leq 0.2$ 
with a fixed $t_{b}/t_{a}=0.1$,
which covers three typical shapes of the Fermi surface 
(Fig.~\ref{fig:03}).
In actual $\mbox{(TMTSF)}_{2}X$ salts,
$|t^{\prime}|$ may be similar to or even greater than $t_{b}$
for some anion $X$ at low temperatures
\cite{Ducasse,Yamaji}.
The chemical potential $\mu$ is set ($\mu=-1.41t_{a}$)
to make the band quarter-filled as in TMTSF.
\par
In the latter half of the paper we apply 
a magnetic field parallel to the $c$-axis ($\perp a,b$).
Since the penetration depth $\lambda \ge 12000$ \AA
\cite{Le93} in $\mbox{(TMTSF)}_2X$
is much greater than the coherence length
$\xi \sim 500$ \AA $(=150a)$
\cite{Le93,Yoshino},
the vector potential can be taken to be 
$\bm{A}=(0,H\lambda,0)$ \cite{FRS97}.
Thus the quasiparticle momentum $k_{b}$ in the $b$ direction
changes as $k_{b}' = k_{b}+H/(\pi \xi H_{c})$,
where $H_{c}=\phi_{0}/(\pi^{2} \xi \lambda)$ 
($= 3.5 \times 10^{-3}$ T for TMTSF) with $\phi_{0}=h/(2e)$.
\par
For the sample with an edge, a mean-field study is performed 
by introducing a site dependent pair potential.  
The symmetry of the pair potential is best represented 
with the $\bm{d}$ vector as \cite{Sigrist}
$\Delta^{\sigma \sigma^{\prime}}_{\bm{ij}}
= \sum_{q}^{0,x,y,z}{\rm i}d^{q}_{\bm{ij}}
(\hat{\sigma}_{q} \cdot \hat{\sigma}_{y})_{\sigma \sigma^{\prime}}$
with 
$d^{q}_{\bm{ij}} = \frac{V}{2}\sum_{\sigma,\sigma^{\prime}}
(\hat{\sigma}_{q} \cdot
\hat{\sigma}_{y})^{\dagger}_{\sigma^{\prime} \sigma}
   \langle c_{\bm{i}\sigma}
    c_{\bm{j}\sigma^{\prime}} \rangle$,
where 
$\hat{\sigma}_{x}$,$\hat{\sigma}_{y}$,
$\hat{\sigma}_{z}$ are Pauli matrices 
and $\hat{\sigma}_{0}=\bm{1}$.
We consider only $d^{z}_{\bm{ij}}(\equiv F^{p}_{\bm{ij}})$ 
with all the other
$d^{0}_{\bm{ij}}=d^{x}_{\bm{ij}}=d^{y}_{\bm{ij}}=0$ 
for triplet $p$-wave state,
$d^{0}(\equiv F^{d}_{\bm{ij}})$ only for singlet $d$, and 
$d^{y}(\equiv F^{f}_{\bm{ij}})$ only for triplet $f$
\cite{KAA01}.
\par
By solving the mean-field equation 
for a region with $N_{\rm L}(=10^{3})$ sites
in the $a$ direction
and one site in the $b$ direction, 
we have obtained the eigenenergy $E_{\nu}$.
In terms of the eigenenergy $E_{\nu}$ and the 
wave functions $u^{\nu}_{\bm{i}}$, $v^{\nu}_{\bm{i}}$,
the Bogoliubov-de Gennes equation for the surface
with the magnetic field applied field parallel to the 
$c$-axis is given by
\begin{eqnarray}
\label{eq:HF}
    \sum_{\bm{j}}
     \left (
      \begin{array}{cc}
    H_{\bm{ij}} &
    F^{l}_{\bm{ij}} \\
    F^{l*}_{\bm{ij}} &
   -H_{\bm{ij}}^{*} \\
      \end{array}
     \right )
     \left (
      \begin{array}{c}
            u^{\nu}_{\bm{j}} \\
            v^{\nu}_{\bm{j}} \\
      \end{array}
     \right )
  =E_{\nu}
     \left (
      \begin{array}{c}
            u^{\nu}_{\bm{i}} \\
            v^{\nu}_{\bm{i}} \\
      \end{array}
     \right ),
\end{eqnarray}
with
%
$H_{ij}(k_{b}) =
    -\sum_{\pm \hat{a}} [ t_{a}\delta_{j,i\pm \hat{a}}
      + 2t_{b}\cos (k_{b}^{\prime}a) \delta_{ij}
    + t^{\prime}e^{\mp {\rm i}k_{b}^{\prime}a}
     \delta_{j,i\pm \hat{a}}
      ] -\mu\delta_{ij}$.
%
\par
The pair potential is determined self-consistently as
\begin{eqnarray}
F^{l}_{\bm{ij}} = -\frac{V}{2}\sum_{k_{b},\nu}
    u^{\nu}_{\bm{i}} v^{\nu \ast}_{\bm{j}}
    \tanh \left [ \frac{E_{\nu}(k_{b})}{2k_{\rm B}T}
    \right ].
\end{eqnarray}

\par

The ABS is probed by 
the surface density of states calculated with 
the pair potential determined self-consistently.
In order to compare our theory with
scanning tunneling microscopy (STM) experiments,
we assume that the STM tip is metallic
with a constant density of states, 
and that the tunneling occurs only for the site nearest to the tip. 
This has been shown to be valid 
through the study of tunneling conductance of unconventional 
superconductors \cite{TK95}. 
The tunneling conductance spectrum is then given
at low temperatures by 
the normalized surface density of states \cite{TK95},
\begin{eqnarray}
   \rho(E) &=&
   \frac{
   \displaystyle{
   \int ^{\infty}_{-\infty}{\rm d}\omega \rho_{\rm S}(\omega)
{\rm sech}^{2}\left ( \frac{\omega + E}{2k_{\rm B}T} \right )}}
   {
   \displaystyle{
   \int ^{\infty}_{-\infty}{\rm d}\omega \rho_{{\rm N}}(\omega)
{\rm sech}^{2}\left ( \frac{\omega - 2\Delta_{l}}{2k_{\rm B}T} \right )}},
\\
\rho_{\rm S}(\omega) &=&
\sum_{k_{b},\nu} \left [|u^{\nu}_{1}|^{2}
\delta(\omega - E_{\nu})
+ |v^{\nu}_{1}|^{2}
\delta(\omega + E_{\nu}) \right ].
\end{eqnarray}
Here $\rho_{\rm S}(\omega)$ denotes the 
surface density of states 
for the superconducting state while $\rho_{\rm N}(\omega)$
the bulk density of states in the normal state.
\par
%
\begin{figure}[htb]
\begin{center}
\scalebox{0.40}{
\includegraphics{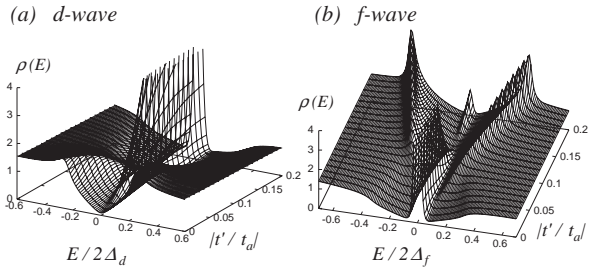}}
\caption{
The surface density of states
at the surface normal to the $a$-axis
in $H=0$
for (a) $d$-wave and (b) $f$-wave.
\label{fig:02}}
\end{center}
\end{figure}
%
Let us first examine the case of zero magnetic field. 
The shape of the gap in the surface density of states 
displayed in Fig.~\ref{fig:02} is 
U-shaped for $p$-wave, while V-shaped for $d$ and $f$.  
This is because the nodal lines (displayed in 
Fig.~\ref{fig:03} by the dashed lines)
in the pair potential intersect the Fermi surface 
for $d$ and $f$, while they do not for $p$-wave, 
as noted in Ref.~\onlinecite{Tanu02} for the case of $t'=0$.
If we turn to the ZEP,
$p$ and $f$-waves have peak \cite{Sengupta,Tanu02}, 
which is due to the fact that 
incident and reflected quasiparticles normal to the surface 
feel opposite signs of the pair potential, which 
results in a formation of the ABS
(whereas oblique incidence is required for $d$-wave).
The situation does not change when $t^{\prime}$
is turned on because the nodes of the $p$-wave lies
away from the Fermi surface, so the warping of 
the Fermi surface has little effect.
\par
%
%
On the other hand, the situation is not so simple
for $d$ and $f$-waves,
where the nodes of the pair potential
intersect the Fermi surface.
For $t^{\prime}=0$, the ZEP appears for $f$,
and does not for $d$ as mentioned in Refs.
\onlinecite{Sengupta} and \onlinecite{Tanu02}.
This is because when the Fermi surface is symmetric 
against $k_a\leftrightarrow -k_a$, 
injected and reflected quasiparticles
always feel opposite (same) signs of the pair potential
for $f(d)$-wave [see Fig.~\ref{fig:03}(a)].
This time the situation does change when we turn on $t^\prime$, 
which makes the Fermi surface
asymmetric against $k_a\leftrightarrow -k_a$,
so that 
some of the Andreev reflection [the blue area in Fig.~\ref{fig:03}(b), 
left panel] lead to ZEP also for d-wave as seen in Fig.~\ref{fig:02}.
For $|t'/t_a|=0.15$ in particular, the warping of the 
Fermi surface is such that injected and reflected quasiparticles 
mostly feel opposite (same) signs 
for $d(f)$ [see Fig.~\ref{fig:03}(c)], 
so the situation for the ZEP is {\it inverted} as seen in
Fig.~\ref{fig:02}.
Consequently, the surface density of states for 
$d$- and $f$-wave pairings look similar (having V-shaped gap + ZEP) 
at around $|t'/t_a|\sim 0.1$, so that it should be difficult 
to distinguish the two by tunneling spectroscopy.
\par
\begin{figure}[htb]
\begin{center}
\scalebox{0.35}{
\includegraphics{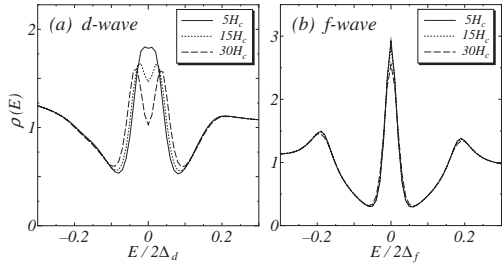}}
\caption{
The surface density of states
normal to $a$ axis  for $|t^{\prime}/t_a|=0.08$
in the presence of the magnetic field for the 
(a) $d$-wave and (b) $f$-wave.
\label{fig:04}}
\end{center}
\end{figure}

This is where the {\it in situ} control of the 
tunneling by magnetic field comes in.
Namely, we now apply magnetic fields
along the $c$-axis.
In Fig.~\ref{fig:04}, we show the magnetic field
dependence of the surface density of states for 
$d$ and $f$-waves for $t'=-0.08t_{a}$.
The ZEP for the $d$-wave is seen to {\it split}
with the magnetic field,
while that for $f$ (or $p$; not shown) does not.
The magnitude of applied magnetic field here is 
$H < 30 H_c$, where $30H_c$ roughly corresponds to 
$H_{c2}^{c}(\sim 0.1$ T), 
i.e., the upper critical field parallel to $c$-axis,
of $\mbox{(TMTSF)}_2\mbox{PF}_6$
\cite{Lee01}.
\par
The result can be interpreted as follows.
For $d$-wave, the ZEP is mainly formed 
by the quasiparticles having $k_{b}\sim \pi/2$,
where 
$v_{{\rm F}b}(k_a)$ and $v_{{\rm F}b}(-k_a)$ have the same sign
[see Fig.~\ref{fig:03}(b), left panel].
This situation is similar to those studied previously
\cite{FRS97,Covin97,Aprili,Krupke}, 
where the ZEP is strongly affected by the Doppler shift.
Namely, the magnetic field gives in this case the 
injected and reflected quasiparticles additional phase shifts 
with the same sign, 
which degrades the constructive interference 
for the formation of ZEP. 
By contrast, for $p$ and $f$-waves quasiparticles
with $k_b=0$ 
mainly contribute to the formation of the zero-energy states
\cite{Buch,Kusakabe}.
For $k_b=0$, $v_{{\rm F}b}(k_a)$ and $v_{{\rm F}b}(-k_a)$ have 
opposite signs regardless of the value of $t'$
[see Fig.~\ref{fig:03}(b), right panel].
In such a case, the magnetic field 
gives the injected and reflected quasiparticles additional 
phase shifts with {\it opposite} signs, which almost cancel out when 
added, so that the formation of ZEP is barely affected
\cite{Tanaka-new}.
%
%
%
This is physically why we can distinguish $d$ and $f$-waves through 
the appearance/disappearance of ZEP splitting 
in the presence of magnetic field. 
\par
%
Although we have here exemplified the tunneling spectra for 
the quasi-1D model, the physics involved 
(i.e., the signs of the pair potential and 
the signs of $v_{\rm F}$ for the incident and 
reflected particles) is quite general, so that 
the principle for distinguishing various singlet and triplet 
pairing symmetries should apply for other, non-cubic 
(or non-tetragonal) systems.
\par
%
Y.T. would like to thank K. Maki and K. Yamaji
for pointing out the possible importance of
the realistic shape of the Fermi surface.
He also acknowledges a financial support
of Japan Society for the Promotion
of Science  for Young Scientists.
This work was in part supported by
the Core Research for
Evolutional Science and Technology (CREST)
of the Japan Science and Technology Corporation.
The computations were performed
at the Supercomputer Center of 
Institute for Solid State Physics,
and the Computer Center in the 
University of Tokyo.
%


\begin{thebibliography}{99}
%
\bibitem{Buch}
L.J. Buchholtz and G. Zwicknagl, Phys. Rev. B 
{\bf 23}, 5788 (1981). 

\bibitem{Hu}
C.R. Hu,
Phys. Rev. Lett. {\bf 72}, 1526 (1994).
%
\bibitem{TK95}
Y. Tanaka and S. Kashiwaya,
Phys. Rev. Lett. {\bf 74}, 3451 (1995);
S. Kashiwaya and Y. Tanaka,
Rep. Prog. Phys. {\bf 63}, 1641 (2000);
T. L\"{o}fwander, V. S. Shumeiko and G. Wendin, 
Supercond. Sci. Technol. {\bf 14}, R53 (2001). 
%
\bibitem{FRS97}
M. Fogelstr\"{o}m, D. Rainer, and J.A. Sauls,
Phys. Rev. Lett. {\bf 79}, 281 (1997).
%
\bibitem{Honer98}
C. Honerkamp and M. Sigrist,
J. Low Temp. Phys. {\bf 111}, 895 (1998).
%
\bibitem{Boss}
M. Yamashiro, $et$ $al$.,
Phys. Rev. B {\bf 56}, 7847 (1997)
%
\bibitem{GXL88}
J. Geerk, 
X.X. Xi, and G. Linker, 
Z. Phys. B. {\bf 73}, 329 (1988).
%
\bibitem{Covin97}
M. Covington, 
$et$ $al$.,  
Phys. Rev. Lett. {\bf 79}, 277 (1997).
%
\bibitem{Wei}
J.Y.T. Wei, 
$et$ $al.$,
Phys. Rev. Lett. {\bf 81}, 2542 (1998). 
%
\bibitem{Alff}
L. Alff, $et$ $al.$,
Phys. Rev. B {\bf 55}, (1997) R14757.
%
\bibitem{Laube}
F. Laube, 
$et$ $al.$,
Phys. Rev. Lett. {\bf 84}, 1595 (2000).
%
\bibitem{Mao}
Z.Q. Mao, 
$et$ $al.$,
Phys. Rev. Lett. {\bf 87}, 037003 (2001).
%
\bibitem{Walti}
Ch. W\"{a}lti, 
$et$ $al.$, 
Phys. Rev. Lett. {\bf 84}, 5616 (2000).
%
\bibitem{Arai}
Tunneling experiment on  (TMTSF)$_2$ClO$_4$ has recently 
been performed by T. Arai, $et$ $al.$,
to be published in Synthetic Metals.
%
\bibitem{Jerome}
D. J\'{e}rome, $et$ $al.$, 
J. Phys. Lett. (France) {\bf 41}, L92 (1980).
%
\bibitem{Bechgaard}
K. Bechgaard,
$et$ $al.$,
Phys. Rev. Lett. {\bf 46}, 852 (1981).
%
\bibitem{Lee01}
I.J. Lee, $et$ $al.$,
Phys. Rev. Lett. {\bf 78}, 3555 (1997);
Phys. Rev. B {\bf 62}, R14669 (2000).
%
\bibitem{Lee02}
I.J. Lee, $et$ $al.$,
Phys. Rev. Lett. {\bf 88}, 017004 (2002).
%
\bibitem{Takigawa}
M. Takigawa, H. Yasuoka, and G. Saito,
J. Phys. Soc. Jpn. {\bf 56}, 873 (1987).
%
\bibitem{BB97}
S. Belin and K. Behnia,
Phys. Rev. Lett. {\bf 79}, 2125 (1997).
%
\bibitem{Abrikosov}
A.A. Abrikosov,
J. Low Temp. Phys. {\bf 53}, 359 (1983).
%
\bibitem{HF87}
Y. Hasegawa and H. Fukuyama,
J. Phys. Soc. Jpn. {\bf 56}, 877 (1987).
%
\bibitem{Lebed}
A.G. Lebed,
Phys. Rev. B {\bf 59}, R721 (1999);
A.G. Lebed, K. Machida, and M. Ozaki,
$ibid.$ {\bf 62}, R795 (2000).
%
\bibitem{GE80}
R.L. Greene and E.M. Engler,
Phys. Rev. Lett. {\bf 45}, 1587 (1980).
%
\bibitem{Shima01}
H. Shimahara,
J. Phys. Soc. Jpn. {\bf 58}, 1735 (1989).
%
\bibitem{KA99}
K. Kuroki and H. Aoki,
Phys. Rev. B {\bf 60}, 3060 (1999).
%
\bibitem{KK99}
H. Kino and H. Kontani,
J. Low. Temp. Phys. {\bf 117}, 317 (1999).
%
\bibitem{KAA01}
K. Kuroki, R. Arita, and H. Aoki,
Phys. Rev. B {\bf 63}, 094509 (2001).
%
\bibitem{Sengupta}
K. Sengupta, $et$ $al.$,
Phys. Rev. B {\bf 63}, 144531 (2001).
%
\bibitem{Tanu02}
Y. Tanuma, $et$ $al.$,
Phys. Rev. B {\bf 64}, 214510 (2001).
%
\bibitem{ZEP}
We use the term ZEP for the peak in the surface 
density of states to distinguish it from the 
conductance peak (ZBCP).
%
\bibitem{Melo}
Quite recently,
the {\it bulk} quasiparticle density of states
of $\mbox{(TMTSF)}_{2}X$ has been 
examined for various pairing symmetries in 
R.D. Ducan, R.W. Cerng and A.R. Sa de Melo,
unpublished [cond-mat/0203570].
%
\bibitem{Aprili}
M. Aprili, E. Badica and L.H. Greene, 
Phys. Rev. Lett. {\bf 83}, 4630 (1999). 
\bibitem{Krupke}
R. Krupke and G. Deutscher, Phys. Rev. Lett. {\bf 83}, 
4634 (1999). 
\bibitem{Ducasse}
L. Ducasse, $et$ $al.$, 
J. Phys. C {\bf 19}, 3805 (1986).
%
\bibitem{Yamaji}
K. Yamaji,
J. Phys. Soc. Jpn. {\bf 55}, 860 (1986).
%
\bibitem{Le93}
L.P. Le, $et$ $al.$, 
Phys. Rev. B {\bf 48}, 7284 (1993).
%
\bibitem{Yoshino}
H. Yoshino, $et$ $al.$, 
J. Phys. Soc. Jpn. {\bf 68}, 3142 (1999).
%
\bibitem{Sigrist}
M. Sigrist and K. Ueda,
Rev. Mod. Phys. {\bf 63}, 239 (1991).
%
\bibitem{Tana02}
Y. Tanaka, $et$ $al.$, 
J. Phys. Soc. Jpn. {\bf 71}, 271 (2002).
%
\bibitem{Kusakabe}
Y. Tanaka, $et$ $al.$, 
Phys. Rev. B {\bf 60}, 6308 (1999). 
%
\bibitem{Tanaka-new}
Y. Tanaka, $et$ $al.$, unpublished.
%
%
\bibitem{comment} 
For $d$-wave with further increase of $|t'|$,
the imaginary part of the pair potential turns out to be 
significant near the surface due to a mixing of $p$-wave,
which results in a broken time-reversal symmetry state,
and makes the splitting of the ZEP much wider.
%
However, we believe that the study of the symmetry mixing 
in the actual $(\mbox{TMTSF})_{2}X$ would require a model 
with {\it spin dependent} electron-electron interaction, so 
we have restricted ourselves to $|t'|\leq 0.08t_a$ 
in the presence of magnetic field, where the symmetry mixing is 
negligible.
\end{thebibliography}
\end{document}